\documentclass[12pt,a4paper]{article}
\usepackage{graphicx}
\usepackage[usenames,dvipsnames]{color}
\usepackage{latexsym}
\usepackage{amsmath}
\usepackage{amssymb}
\usepackage{url}
\usepackage{algorithmicx}
\usepackage[ruled]{algorithm}
\usepackage{algpseudocode}
\usepackage{booktabs}

\newcommand{\figref}[1]{Fig.~\ref{fig:#1}}

\newcommand{\kmaxc}{k_{\max}^{\rm c}}
\newcommand{\kmaxf}{k_{\max}^{\rm f}}

\begin{document}

\title{Cycle and flow trusses in directed networks}

\author{Taro Takaguchi$^{1,2,\dagger,}$\footnote{Present address: National Institute of Information and Communications Technology, 4-2-1 Nukui-Kitamachi, Koganei, Tokyo 184-8795, Japan}, Yuichi Yoshida$^{1,3,\ddagger}$\\
\\
\\
${}^{1}$
National Institute of Informatics,\\
2-1-2 Hitotsubashi, Chiyoda-ku, 101-8430 Tokyo, Japan
\\
\\
${}^{2}$
JST, ERATO, Kawarabayashi Large Graph Project,\\
2-1-2 Hitotsubashi, Chiyoda-ku, 101-8430 Tokyo, Japan
\\
\\
${}^{3}$
Preferred Infrastructure, Inc.,\\
1-6-1 Otemachi, Chiyoda-ku, 100-0004 Tokyo, Japan
\\
\\
${}^\dagger$ t\_takaguchi@nii.ac.jp\\
${}^\ddagger$ yyoshida@nii.ac.jp
}

\setlength{\baselineskip}{0.77cm}

\maketitle

\newpage

\begin{abstract}
\setlength{\baselineskip}{0.77cm}
When we represent real-world systems as networks, the directions of links often convey valuable information.
Finding module structures that respect link directions is one of the most important tasks for analyzing directed networks.
Although many notions of a directed module have been proposed, no consensus has been reached.
This lack of consensus results partly because there might exist distinct types of modules in a single directed network, whereas most previous studies focused on an independent criterion for modules.
To address this issue, we propose a generic notion of the so-called truss structures in directed networks.
Our definition of truss is able to extract two distinct types of trusses, named the cycle truss and the flow truss, from a unified framework.
By applying the method for finding trusses to empirical networks obtained from a wide range of research fields, we find that most real networks contain both cycle and flow trusses.
In addition, the abundance of (and the overlap between) the two types of trusses may be useful to characterize module structures in a wide variety of empirical networks.
Our findings shed light on the importance of simultaneously considering different types of modules in directed networks.
\end{abstract}

\newpage

\section{Introduction}
Analysis methods developed in network science provide us with useful tools for investigating and characterizing the kinds of network structures observed in real-world systems~\cite{Newman2010}.
Standard techniques in network science include characterizing global properties of networks, measuring centralities of nodes and links, and classifying nodes into groups~\cite{Costa2007,Barrat2008}.
Finding relevant subgroups of nodes, often called communities or modules, is a fundamental problem.
This problem is referred to as the community detection problem~\cite{Porter2009,Fortunato2010}, which has been studied in different disciplines including computer science, statistics, and statistical physics.
Although there is no unifying definition of a community, information regarding communities in networks gives us a guide to summarize large-scale networks~\cite{Rosvall2008}, to predict the existence of links~\cite{Clauset2008}, and to reveal functional organization in real networks~\cite{Chen2006,Sohn2011}.

Within the community detection problem for real-world networks, the direction of links plays a crucial role.
Although the majority of previous notions of a community assume undirected networks as their target, several of them are explicitly designed for directed networks.
Examples include extensions of graph conductance~\cite{Chung2005,Michoel2012,Yoshida2016}, a generalization of the modularity function~\cite{Leicht2008}, and the map equation~\cite{Rosvall2008} (see Ref.~\cite{Malliaros2013} for a comprehensive review on the community detection problem in directed networks).
These definitions of communities often successfully detect communities that satisfy their criteria.
Nevertheless, it remains an open problem as to how to choose a notion of modules when we are given directed network data.
For undirected networks, the local density of links within a subgroup of nodes is arguably a suitable criterion for a module, regardless of the details of the definitions~\cite{Fortunato2010}.
In contrast, for directed networks, the directionality of links can alter the module structure, even when we observe the same link density in two subgroups of nodes.
The choice of algorithms crucially depends on what types of modules we expect to find.
In addition, some definitions of communities and their associated algorithms are known to fail to detect certain types of module structure~\cite{Kim2010,Malliaros2013}; the impact of this drawback is not clear until we analyze the network.
Therefore, a generic notion of directed modules is necessary for understanding the nature of module structures in real-world directed networks.

To address this issue, in this paper, we propose notions of module structures in directed networks.
The following observations underlie the core concept of our work.
First, there could be different types of modules within a single directed network.
For example, one part of a network can be an all-to-all connected module, whereas another part can form a layered structure~\cite{Malliaros2013}.
Second, it is not necessary to divide the entire network into modules.
For example, the World Wide Web network is well known to exhibit the so-called bow-tie structure~\cite{Broder2000,Corominas-Murtra2013}.
This fact implies that the different parts of a directed network may not be regarded as modules to an equal degree.
Instead of partition of networks into modules, extraction of modules from the network should be considered.
Previous studies often ignore these observations: they aim at partitioning the entire network into modules based on a single objective function.
Therefore, we propose two distinct types of modules, called the cycle truss and the flow truss, using a unified framework, and an algorithm for finding them.
Our definition of trusses relies on pattern matching and local agglomeration of directed triangles (i.e., connected subgraphs composed of three nodes and three links).

We apply the proposed algorithm for finding trusses to a variety of empirical networks to verify its practicality.
First, we observe that the extracted trusses seem to capture meaningful subgroups of nodes in networks with node label data.
Second, because our method simultaneously detects the two types of modules, we can use them as features for classifying different networks.
Empirical networks obtained from the same categories (e.g., social or biological) tend to show a similar degree of abundance of the two types of trusses.
In addition, the overlap between the two types of trusses captures another kind of similarity between networks in a given category.
Our findings demonstrate the importance of simultaneously considering different types of modules in directed networks to understand the common properties underlying the organization of modules.

\section{Methods}

\subsection{Definitions of cycle and flow trusses}
\begin{figure}
\centering
\includegraphics[width=0.5\hsize]{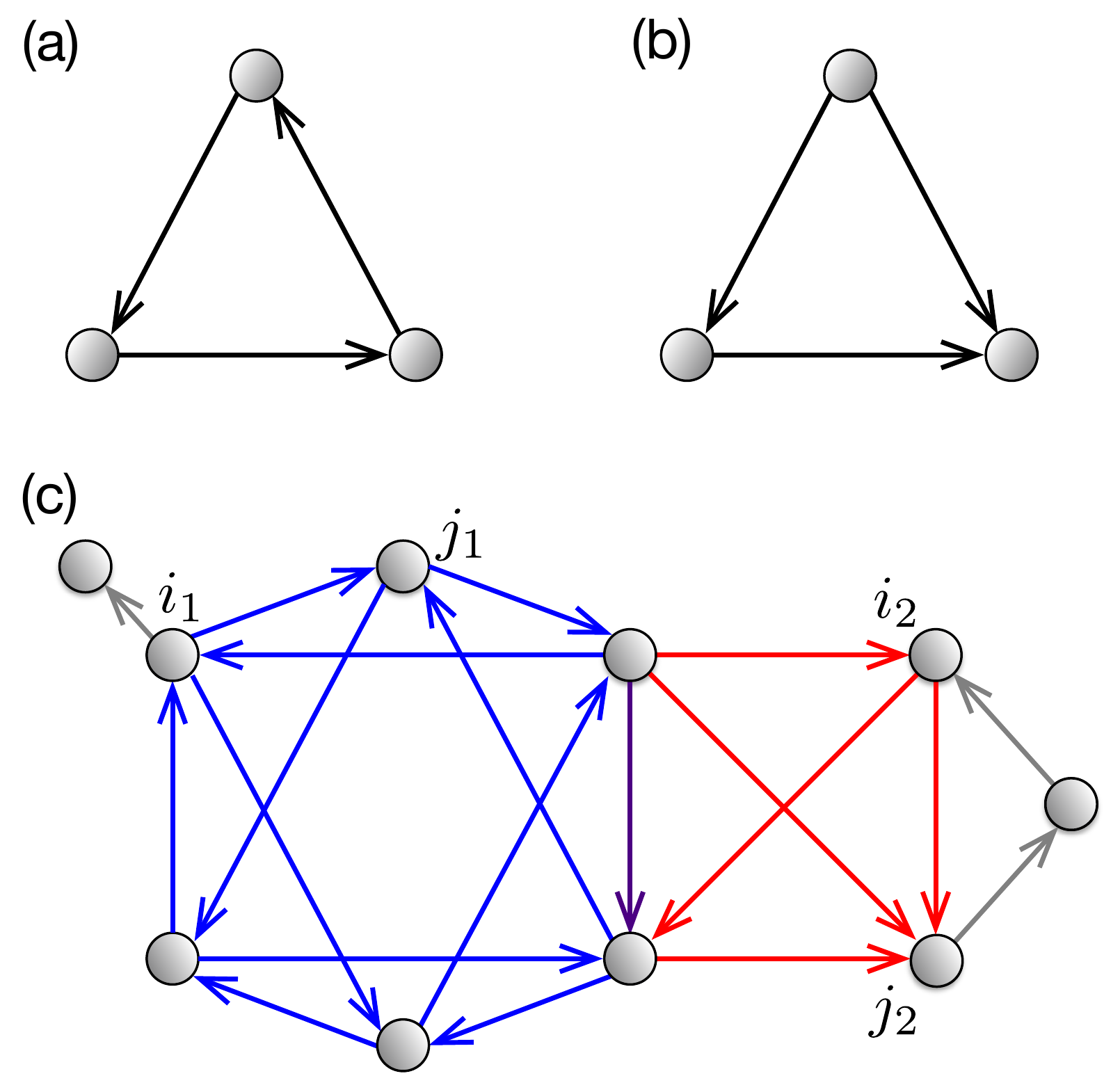}
\caption{Definitions of cycle and flow trusses. \textbf{(a)} a cycle and \textbf{(b)} flow triangles. \textbf{(c)} the cycle (blue) and flow (red) $k=2$-trusses in an example network.
The vertical arrow colored with purple at the center represents the link that belongs to both of the cycle and flow $2$-trusses.}
\label{fig:triangles}
\end{figure}

In this paper, we assume that the focal network is directed and simple, i.e., there are no self-loops and no multiple links between any pair of nodes in the same direction.
Bidirectional connections between two nodes are possible: links from node $i$ to node $j$ and from $j$ to $i$ may coexist.
We also assume that links are unweighted. 
First, we define cycle and flow triangles as the elements of cycle and flow trusses, respectively.
A cycle triangle is a connected subgraph composed of three nodes all of which have out-degree equal to one, namely a directed cycle composed of three nodes (see~\figref{triangles}(a)).
A flow triangle is a connected subgraph composed of three nodes that have out-degrees equal to zero, one, and two (see~\figref{triangles}(b)).
A flow triangle is also called a feed-forward loop~\cite{Milo2002,Alon2007}.
Next, we define the cycle and flow trusses by generalizing the $k$-truss, originally defined for undirected networks~\cite{Cohen2008}.
A cycle (flow) $k$-truss is defined as a maximal connected subgraph of a network in which every link is involved in at least $k$ cycle (flow) triangles within the subgraph~\cite{note_truss_def} (see~\figref{triangles}(c) for an example).
Free parameter $k$ takes a nonnegative integer value in $0 \leq k \leq d_{\max}-1$, where$d_{\max}$ denotes the maximum node degree in the network, and $k$ controls the extent to which triangles are overlapped within a truss.
It should be noted that a cycle (flow) truss may contain flow (cycle) triangles.
There are several relationships between the cycle and flow trusses and other notions of directed subgraphs presented in previous studies, and we will describe this point in Discussion.

The definitions of the cycle and flow $k$-trusses satisfy the requirements described in Introduction, as we can see in the example shown in~\figref{triangles}(c).
To be more precise, these definitions enable us to find two distinct types of modules using the unified framework.
In addition, this method extracts modules from a network, instead of partitioning the entire network into modules.
The algorithm for finding cycle and flow $k$-trusses in a given network is a modified version of that for undirected truss~\cite{Wang2012}.
The details of the algorithm are presented with pseudo-codes in Supplementary Materials (SM).

The definitions of the trusses lead to their basic properties as follows:
First, there can be multiple cycle (flow) $k$-trusses in a network.
Second, cycle and flow $(k+a)$-trusses are, if exist, subgraphs of cycle and flow $k$-trusses, respectively $(a = 1,2, \ldots)$.
Third, the node sets of two $k$-trusses of the same type can overlap, but their link sets must be disjoint because of the maximal property in the definition.
Fourth, the complete graph with $k$ nodes in which all node pairs are connected for both directions is both a cycle $(k-2)$-truss and a flow $3(k-2)$-truss at the same time $(k \geq 3)$.

We assign a link from nodes $i$ to $j$ with the truss number $k_{i \to j}$ defined by
\begin{equation}
k_{i \to j} \equiv \max \left\{ k \mid (i \to j) \in E_k \right\},
\end{equation}
where $E_k$ is the set of links involved in $k$-trusses.
We denote the truss number for cycle and flow trusses by $k^{\rm c}_{i \to j}$ and $k^{\rm f}_{i \to j}$, respectively.
We use the superscripts `c' and `f' to represent the variables related to the cycle and flow trusses throughout this paper.
The truss number indicates the extent of agglomeration of triangles around a link.
For example, in the network shown in~\figref{triangles}(c), link $(i_1 \to j_1)$ has $(k^{\rm c}, k^{\rm f}) = (2, 0)$ and  link $(i_2 \to j_2)$ has $(k^{\rm c}, k^{\rm f}) = (1, 2)$.
We are also interested in the maximum values of $k^{\rm c}_{i \to j}$ and $k^{\rm f}_{i \to j}$ over all the links.
We call these values the maximum truss numbers; they are denoted by $\kmaxc$ and $\kmaxf$ for cycle and flow trusses, respectively.
For the network shown in~\figref{triangles}(c), $\kmaxc = \kmaxf = 2$ holds.

\section{Results}
\subsection{Trusses in empirical networks}
We apply the proposed method to empirical network data that are assigned with predefined node labels so as to demonstrate that cycle and flow trusses can extract meaningful modules.
For this purpose, we use two networks obtained from different fields: the neural network of \textit{Caenorhabditis elegans} (\textit{C.~elegans})~\cite{Varshney2011} and the network between words collected via word association experiments, so-called the Edinburgh Associative Thesaurus (EAT)~\cite{Kiss1973}.

\subsubsection{Neural network of \textit{Caenorhabditis elegans}}\label{sec:cele}
The \textit{C.~elegans} neural network comprises $279$ nodes, which correspond to neurons, and $2,990$ links between the nodes.
A chemical synapse between two neurons is represented as a directed link and an electrical junction as two directed links in both directions between the node pair~\cite{Varshney2011}.
Each neuron is assigned with a unique name and additional information such as soma positions in the worm's body and functional categories (i.e., sensory neuron, interneuron, or motor neuron), which allows us to interpret the neuronal functions of the extracted trusses.

\begin{figure}
\centering
\includegraphics[width=0.8\hsize]{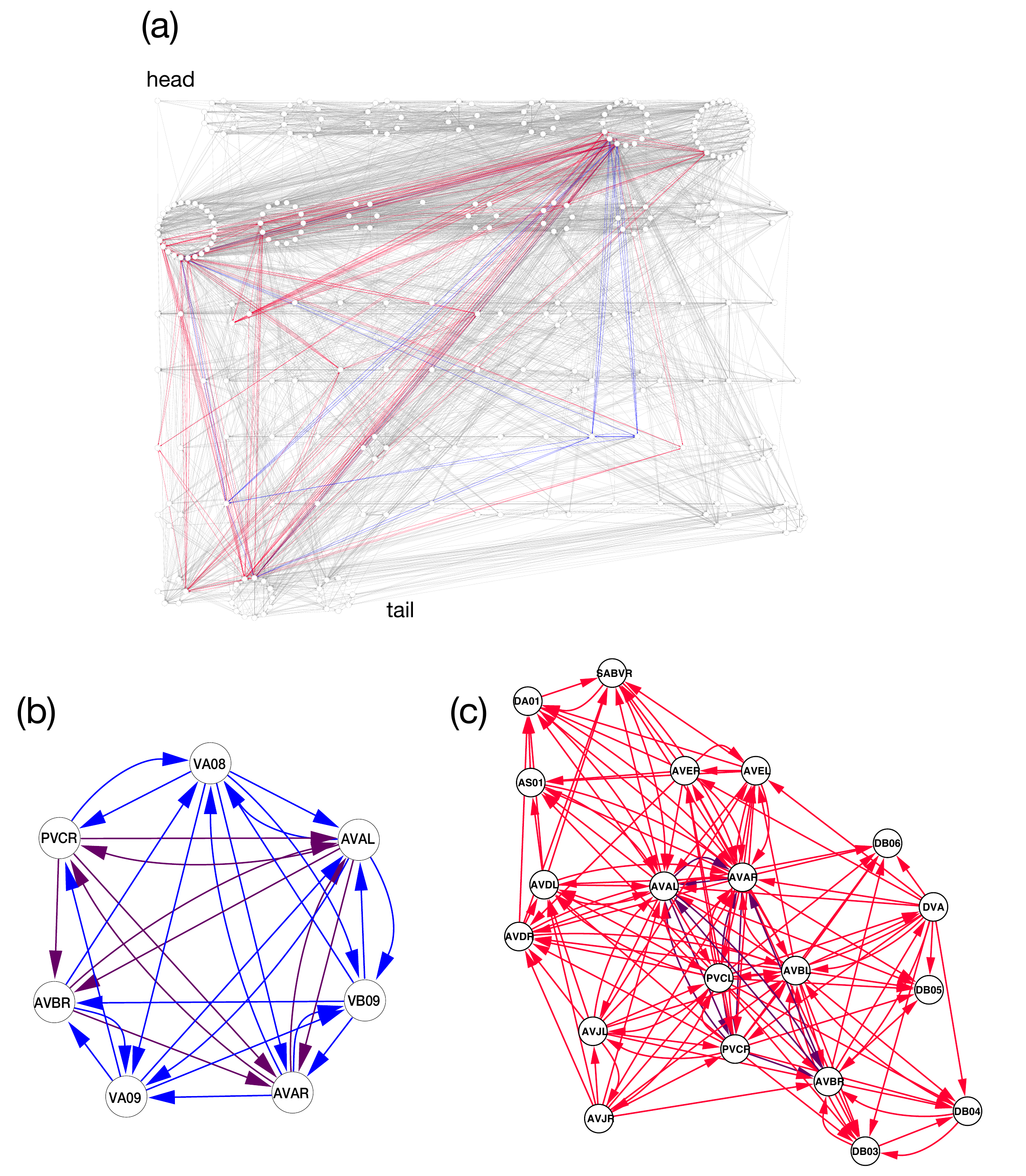}
\caption{
Cycle and flow trusses in the \textit{C.~elegans} neural network.
We set $k$ to $\kmaxc = 3$ and $\kmaxf = 9$ for the cycle and flow trusses, respectively.
The links are colored with blue (in the cycle $3$-truss), red (in the flow $9$-truss), purple (in both), and gray (remainder).
\textbf{(a)} The whole picture of the \textit{C.~elegans} neural network. The nodes are ordered according to the soma position in the worm body (head to tail from left to right and from top to bottom).
A group of nodes composing a circle have the same position.
\textbf{(b)} the cycle $3$- and \textbf{(c)} the flow $9$-trusses. The node labels indicate the names of neurons.
}
\label{fig:demo_cele}
\end{figure}

In~\figref{demo_cele}, the resulting $k$-trusses are depicted.
We focus on the cycle $\kmaxc = 3$- and flow $\kmaxf = 9$-trusses, which are the most cohesive trusses in the network.
In this case, the cycle $3$-truss is a single strongly connected component, and the flow $9$-trusses is a single weakly connected component (i.e., any pair of nodes in each of the trusses is connected if we discard the link direction).
When we map the cycle and flow trusses in the entire network (\figref{demo_cele}(a)), neither of these trusses is localized in any particular part of the worm body, but instead, spans almost the entire range of the body (from head to tail).
We can see that a small number of nodes bridges most of the triangles in the trusses.
The cycle $3$-truss (\figref{demo_cele}(b)) consists of four command interneurons relevant for locomotion (AVAL, AVAR, AVBR, and PVCR) and three motor neurons in the ventral cord of the worm (VA08, VA09, and VB09).
Here, we follow the description of each neuron in Ref.~\cite{Altun2013}.
These seven neurons in the cycle $3$-truss are tightly connected to each other; however, certain node pairs have a link in only one direction.
On the other hand, the flow $9$-truss (\figref{demo_cele}(c)) consists of 13 interneurons relevant to locomotion (AVAL, AVAR, AVBL, AVBR, AVDL, AVDR,  AVEL, AVER, AVJL, AVJR, PVCL, PVCR, and SABVR) and seven motor neurons in the dorsal cord of the worm (DA1, DB03 to 06, and DVA), except for one in the ventral cord (AS01).
Although we need further explanation by biological experts as to why these neurons constitute the cycle and flow trusses, the trusses seem to represent some functional modules of neurons.
The cycle and flow trusses overlap with each other and have four command interneurons AVAL, AVAR, AVBR, and PVCR in common.
This fact arises logically, as the command interneurons related to locomotion should play a central role in mediating the motor neurons in the ventral and dorsal cords~\cite{Varshney2011}.
This example of the \textit{C.~elegans} neural network demonstrates the ability of our proposed method to extract different types of cohesive modules from networks.
In addition, the overlap between the two types of trusses can shed light on the importance of nodes that bridge different modules.

\subsubsection{Word network of the Edinburgh Associative Thesaurus}
Our second example is the graph representation of the Edinburgh Associative Thesaurus (EAT), which was collected through word association experiments with subjects~\cite{Kiss1973}.
A directed link from nodes (i.e., words) $i$ to $j$ represents the associative relationship between the two: for subjects, word~$j$ comes to mind when they are shown word~$i$ as a stimulus.
After the aggregation of the results of word association experiments for many subjects with different stimuli, the EAT network contains $23,219$ nodes and $325,029$ links between them.

\begin{figure}
\centering
\includegraphics[width=0.9\hsize]{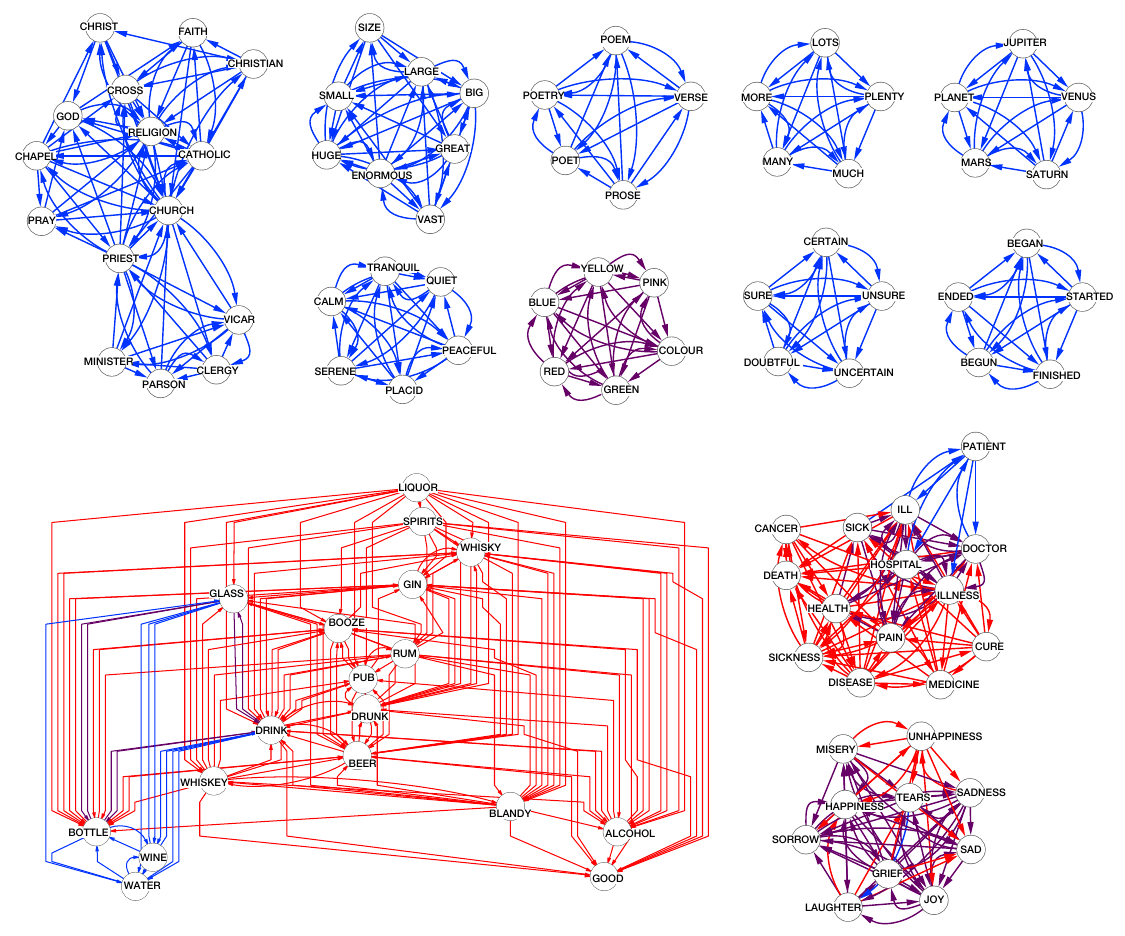}
\caption{
Cycle and flow trusses in the EAT network.
We set $k$ to $\kmaxc = 2$ and $\kmaxf = 8$ for the cycle and flow trusses, respectively.
The links are colored with blue (in the cycle $2$-trusses), red (in the flow $8$-trusses), and purple (in both).
The node labels indicate the corresponding words.}
\label{fig:demo_eat}
\end{figure}

In~\figref{demo_eat}, the resulting $k$-trusses are depicted.
We show the cycle $\kmaxc = 2$- and flow $\kmaxf = 8$-trusses.
Unlike the results of the \textit{C.~elegans} neural network (\figref{demo_cele}),
there are multiple disjoint cycle and flow trusses with $\kmaxc = 2$ and $\kmaxf = 8$.
We can see that each of the trusses consists of words related to a topic, for example, religion, emotion, health, and poem.
All six of the cycle $2$-trusses composed of five nodes are fully connected, in which all node pairs have links in both directions.
Each of the cycle $2$-trusses related to emotion, health, and color strongly overlap with one of the flow $8$-trusses (top center and bottom right of \figref{demo_eat}).
Only the flow $8$-truss related to liquor, the largest one (bottom left of \figref{demo_eat}), is less overlapped with the cycle trusses than the other flow $8$-trusses are.
We can intuitively explain the difference between the cycle and flow trusses in this example network as follows.
The words constituting a cycle truss have an equal relationship with each other such that the experimental subjects tend to recall all words based on each word.
By contrast, the words constituting a flow truss have a hierarchical relationship such that some words remind the subjects of other words but the converse rarely occurs.
If we discard the link direction, we cannot distinguish the modules of the cycle and flow trusses.
Therefore, this example demonstrates that the link direction plays an important role in finding modules.

\subsection{Classification of networks based on truss number distributions}
We can use the truss number statistics to classify various networks.\footnote{Sources of the network data (date accessed: 1st March 2016). The airport, communication, following, and software networks (\url{http://konect.uni-koblenz.de/});
the USairport500 network (\url{http://toreopsahl.com/datasets/\#usairports});
the circuit networks and the word networks (\url{http://www.weizmann.ac.il/mcb/UriAlon/download/collection-complex-networks});
the allcites network (\url{http://fowler.ucsd.edu/judicial.htm});
the cit-HepPh, cit-HepTh, social, slashdot-0902, twitter\_combined, wiki-Vote, P2P, and the web networks (\url{http://snap.stanford.edu/data/});
the food webs and the Edinburgh Associative Thesaurus (\url{http://vlado.fmf.uni-lj.si/pub/networks/data/});
the gene regulatory networks (\url{http://info.gersteinlab.org/Hierarchy});
the \textit{Caenorhabditis elegans (C.~elegans)} neural network (\url{http://www.wormatlas.org/neuronalwiring.html});
the brain connectivity networks (\url{https://sites.google.com/site/bctnet/datasets});
the mac95 network (\url{http://www.biological-networks.org/?page_id=25});
the polblog network (\url{http://www-personal.umich.edu/~mejn/netdata/});
the metabolic networks (personal communication with Kazuhiro Takemoto, 2015)}
In the following, we will demonstrate the classification of networks based on the truss number statistics in two ways. First, we will separately quantify the abundance of cycle and flow trussses and use them as two features. Second, we will quantify the overlap between the cycle and flow trusses with large $k$ values.

\begin{figure}
\centering
\includegraphics[width=\hsize]{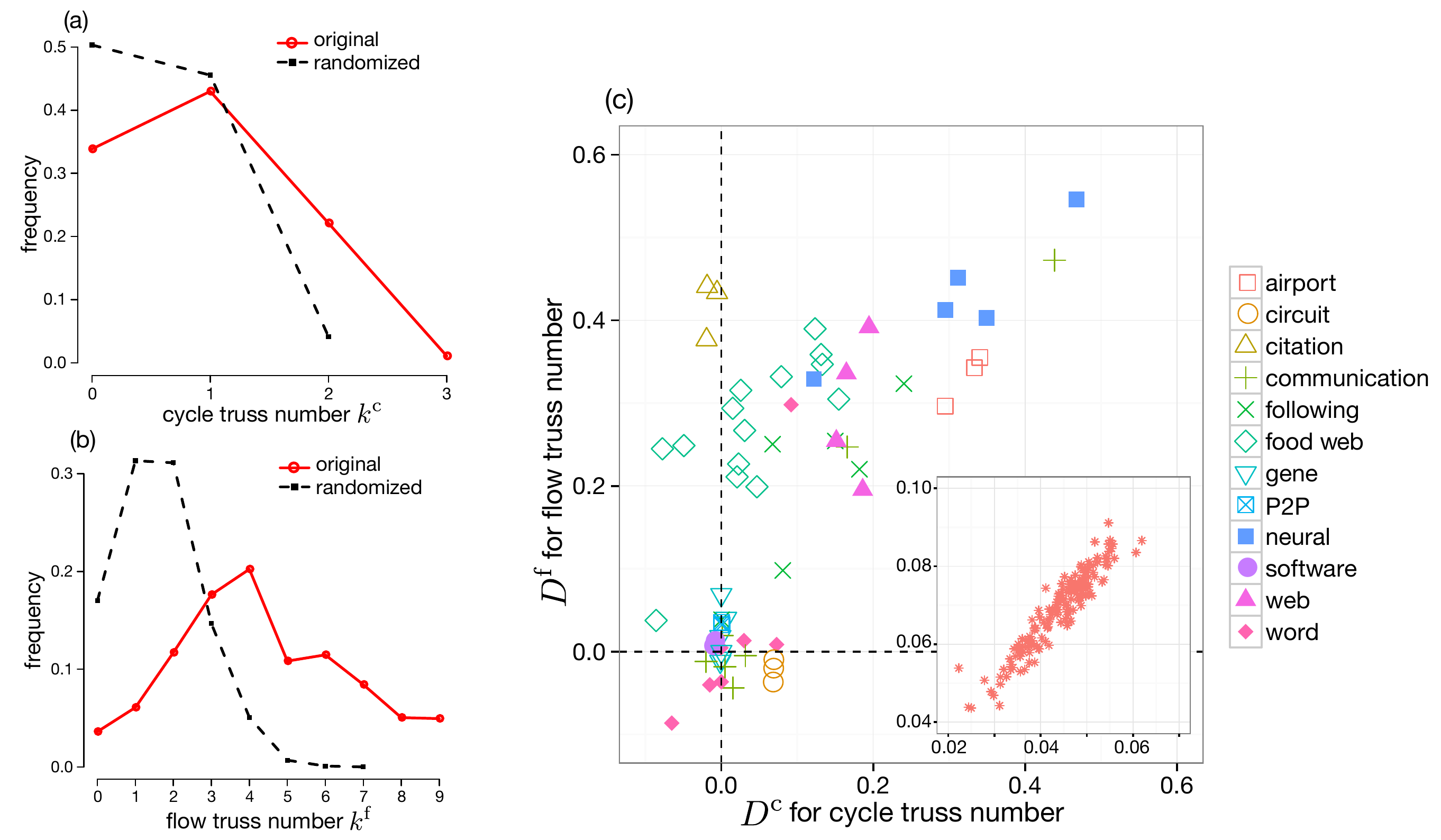}
\caption{
Distributions of the truss numbers and the $D$ measure. Distributions of \textbf{(a)} $k^{\rm c}$ and \textbf{(b)} $k^{\rm f}$ for the original network and $100$ randomized networks of the \textit{C.~elegans} neural network.
\textbf{(c)} Scatter plot of $D^{\rm c}$ and $D^{\rm f}$ for (main panel) empirical networks of the 12 categories and for (inset) the metabolic networks.}
\label{fig:truss_number_D}
\end{figure}

Intuitively, a network is more cycle (flow) truss oriented if the links tend to have larger cycle (flow) truss numbers.
Note that a large truss number implies the agglomeration (and abundance) of triangles.
To quantify how much a network is truss oriented, we define a measure $D$ as
\begin{equation}
D \equiv \frac{1}{K} \sum_{k=0}^{K} \left( F_{\rm rand}(k) - F_{\rm orig}(k) \right),
\label{eq:D}
\end{equation}
where $F(k)$ is the complementary cumulative distribution of truss numbers defined by $F(k) \equiv \sum_{k^\prime=0}^k f(k^\prime)$ and $f(k^\prime)$ (in the sum) is the frequency distribution of the truss number.
In Eq.~\ref{eq:D}, the subscripts ``orig'' and ``rand'' represent the distributions for the original and randomized networks, respectively.
We randomize the original network by rewiring links in a uniformly random manner while retaining the in- and out-degrees of all nodes (i.e., the configuration model for directed networks~\cite{Newman2001}).
The range of the sum over $k$ is determined by $K$. Here, we choose $K \equiv \min \left\{ k \mid F_{\rm orig}(k) > 0.9 \ \wedge \  F_{\rm rand}(k) > 0.9 \right\}$.
We do not assume $K = k_{\max}$, because $k_{\max}$ might be sensitive to noise in the network data.
The measure $D$ takes a value in $[-1, 1]$; a large positive value of $D$ represents that the links in the original network tend to have a larger truss number than those in the randomized networks.
We denote the measure $D$ for the cycle and flow truss numbers by $D^{\rm c}$ and $D^{\rm f}$, respectively.
In Figs.~\ref{fig:truss_number_D}(a) and \ref{fig:truss_number_D}(b), we plot the truss number distributions $f^{\rm c}(k)$ and $f^{\rm f}(k)$ of the \textit{C.~elegans} neural network and the randomized networks.
The $k_{\max}^{\rm c}$ and $k_{\max}^{\rm f}$ values are larger for the original network than those for the randomized networks.
The proportions of links with large $k$ values are also larger for the original than the randomized networks.
For this network, we obtain $(D^{\rm c}, D^{\rm f}) = (0.122, 0.329)$.
Therefore, the \textit{C.~elegans} network is inclined to have more flow trusses than cycle trusses, which agrees well with our intuition based on Figs.~\ref{fig:truss_number_D}(a) and \ref{fig:truss_number_D}(b).

The measure $D$ allows us to compare various networks of different sizes in terms of truss tendency.
In \figref{truss_number_D}(c), we show the scatter plot of $D^{\rm c}$ and $D^{\rm f}$ for the empirical networks.
As we can see, the networks of certain categories such as airport, circuit, citation, and food webs loosely fall into the similar positions on the plane.
The points are located around the diagonal because networks with larger numbers of triangles tend to have larger $D$ values.
This phenomenon occurs because our randomization procedure does not conserve the number of triangles in the original network; the randomization tends to destroy triangles, and consequently, destroy the truss structure.
This reasoning is supported by the observation shown in Fig.~S1 in SM.
The elements of the first principle component of the plot shown in \figref{truss_number_D}(c) exhibit a strong positive correlation with the clustering coefficient~\cite{Watts1998} after discarding the link directions.
Nevertheless, \figref{truss_number_D}(c) provides the information regarding network topology beyond simply the count of the triangles.
For example, there are several distinguishable classes of network (such as citation and circuit networks) in which either cycle or flow trusses are dominant.
The neural, airport, and web networks contain both types of trusses.
The metabolic networks tend to have few cycle or flow trusses as comparable with those in the randomized networks.
These observations suggest the usefulness of the cycle and flow trusses to characterize different types of directed networks.

The randomization method that we used above generally destroys triangles, and randomization conserving the number of triangles is desirable to distinguish the abundance and agglomeration of triangles. Such a randomization method is proposed in Ref.~\cite{Michoel2011} which conserves the number of focal motif counts, while it is basically a rejection sampling and computationally demanding and feasible only for small networks. We applied this randomization method to the food web networks so as to retain both the number of the cycle triangles and that of the flow triangles, and performed the same set of analysis of the $D$ measure (see Fig.~S2 for the resulting plot). The result indicates that the food web networks are more flow-truss oriented than to be cycle-truss oriented, which is consistent with the conclusion based on the randomization method without conserving the number of triangles (Fig.~\ref{fig:truss_number_D}).

\begin{figure}
\centering
\includegraphics[width=\hsize]{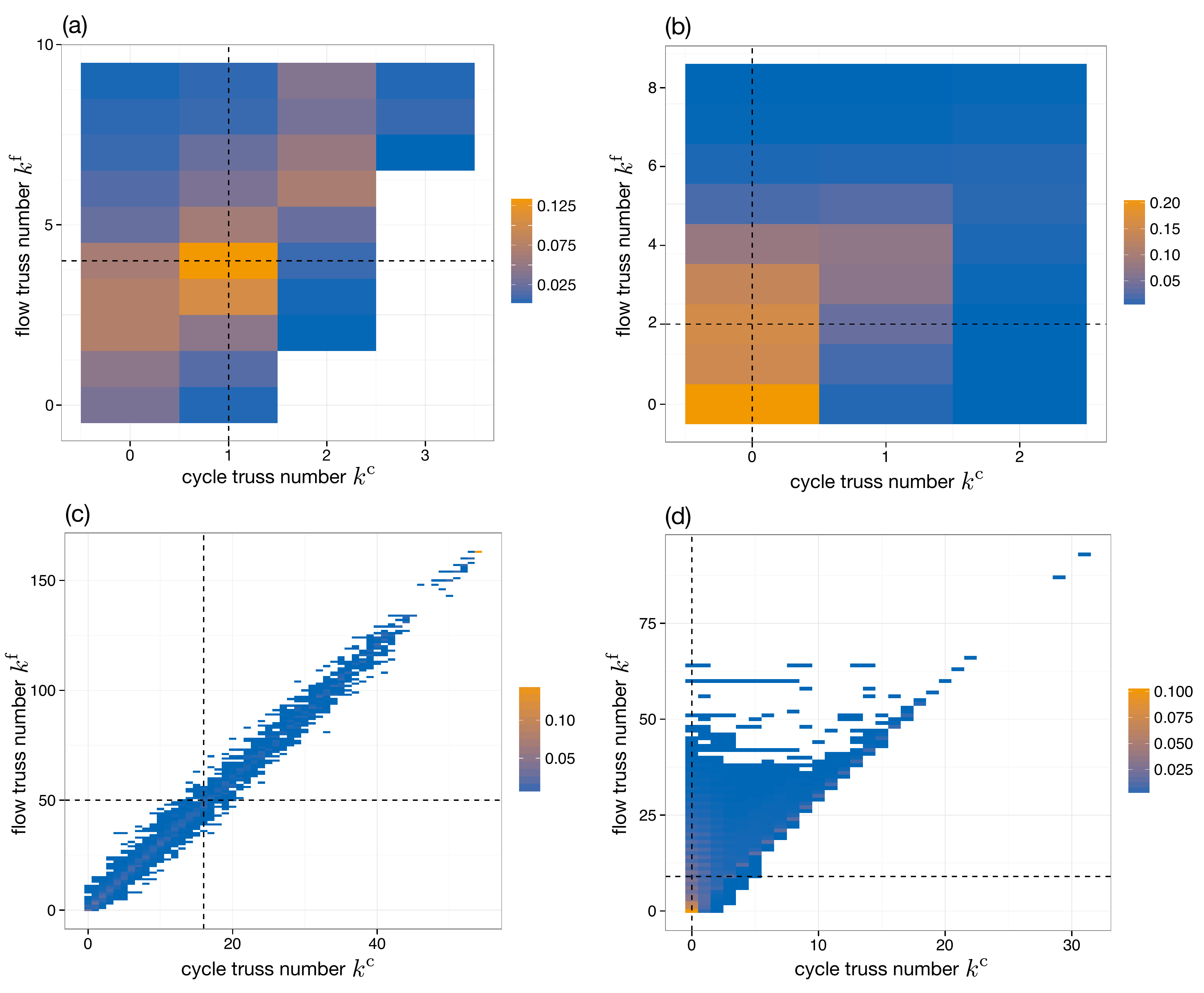}
\caption{
Joint frequency distribution of $\left(k^{\rm c}, k^{\rm f} \right)$. \textbf{(a)} the \textit{C.~elegans} neural network.
\textbf{(b)} the EAT network. \textbf{(c)} the USairport 2010 network. \textbf{(d)} the web-Google network.
The vertical and horizontal dashed lines indicate the $k$ values that gives the median value for $k^{\rm c}$ and $k^{\rm f}$, respectively.
}
\label{fig:joint_dist}
\end{figure}

While we separately considered the properties related to the cycle and flow trusses so far, the overlap between the two types of trusses can be another characteristic of the networks as we observed in the example networks (Figs.~\ref{fig:demo_cele} and~\ref{fig:demo_eat}).
In particular, we are interested in the overlap between highly cohesive cycle and flow trusses with large $k$ values.
To analyze the overlap, we plot the joint frequency distribution of truss numbers $(k^{\rm c}, k^{\rm f})$ for four example networks, i.e., the \textit{C.~elegans} neural network, the EAT network, the USairport 2010 network~\cite{Opsahl2011}, and the web-Google network~\cite{Leskovec2009} (\figref{joint_dist}).
In these plots, a cell at $(k^{\rm c}, k^{\rm f})$ indicates the proportion of links with these truss numbers.
These plots indicate the unique characteristics of the different networks.
In the \textit{C.~elegans} neural network (\figref{joint_dist}(a)), the links with $k^{\rm c} = 3$ have only $7 \leq k^{\rm f} \leq 9 = k_{\max}^{\rm f}$.
This property suggests that the size of the cycle $\kmaxc$-truss is smaller than that of the flow $\kmaxf$-truss and a large section of the cycle truss overlaps with the flow truss, as we observed in~\figref{demo_cele}.
By contrast, for the EAT network (\figref{joint_dist}(b)), the majority of links have relatively small truss numbers, as $k^{\rm c} = 0$ and $0 \leq k^{\rm f} \leq 3$.
Thus, the cohesive cycle and flow trusses are not strongly overlapped.
The plots for the USairport 2010 (\figref{joint_dist}(c)) and web-Google (\figref{joint_dist}(d)) networks look similar, such that we can see the colored cells along the diagonal of slope equal to three.
This observation suggests that in these networks there are complete subgraphs within which node pairs are connected in both directions.
This situation may correspond to airports within local regions and web pages under the same directories.

To quantify the overlap between the cohesive cycle and flow trusses in a network, we define
\begin{equation}
R \equiv
\frac{\left| \left\{ e \in E \mid \left(k_e^{\rm c} > k_{\rm med}^{\rm c} \right) \wedge \left( k_e^{\rm f} > k_{\rm med}^{\rm f} \right) \right\} \right|}
{\left| \left\{ e \in E \mid \left( k_e^{\rm c} > k_{\rm med}^{\rm c}\right) \vee \left( k_e^{\rm f} > k_{\rm med}^{\rm f} \right) \right\} \right|},
\end{equation}
where $E$ is the set of all links and $k_{\rm med}^{\rm c}$ and $k_{\rm med}^{\rm f}$ are the median values of $f^{\rm c}(k)$ and $f^{\rm f}(k)$ (indicated by the dashed lines in~\figref{joint_dist}), respectively.
The measure $R$ characterizes the proportion of links with large $k^{\rm c}$ \emph{and} $k^{\rm f}$ values among those with large $k^{\rm c}$ \emph{or} $k^{\rm f}$ values.
The measure $R$ takes a value in $[0,1]$; a large $R$ value indicates a strong overlap between the cohesive cycle and flow trusses.
For the four networks shown in~\figref{joint_dist}, we obtain $R=0.531$, $0.383$, $0.983$, and $0.540$ for the \textit{C.~elegans} neural network, the EAT network, the USairport 2010 network, and the web-Google network, respectively.

\begin{figure}
\centering
\includegraphics[width=0.8\hsize]{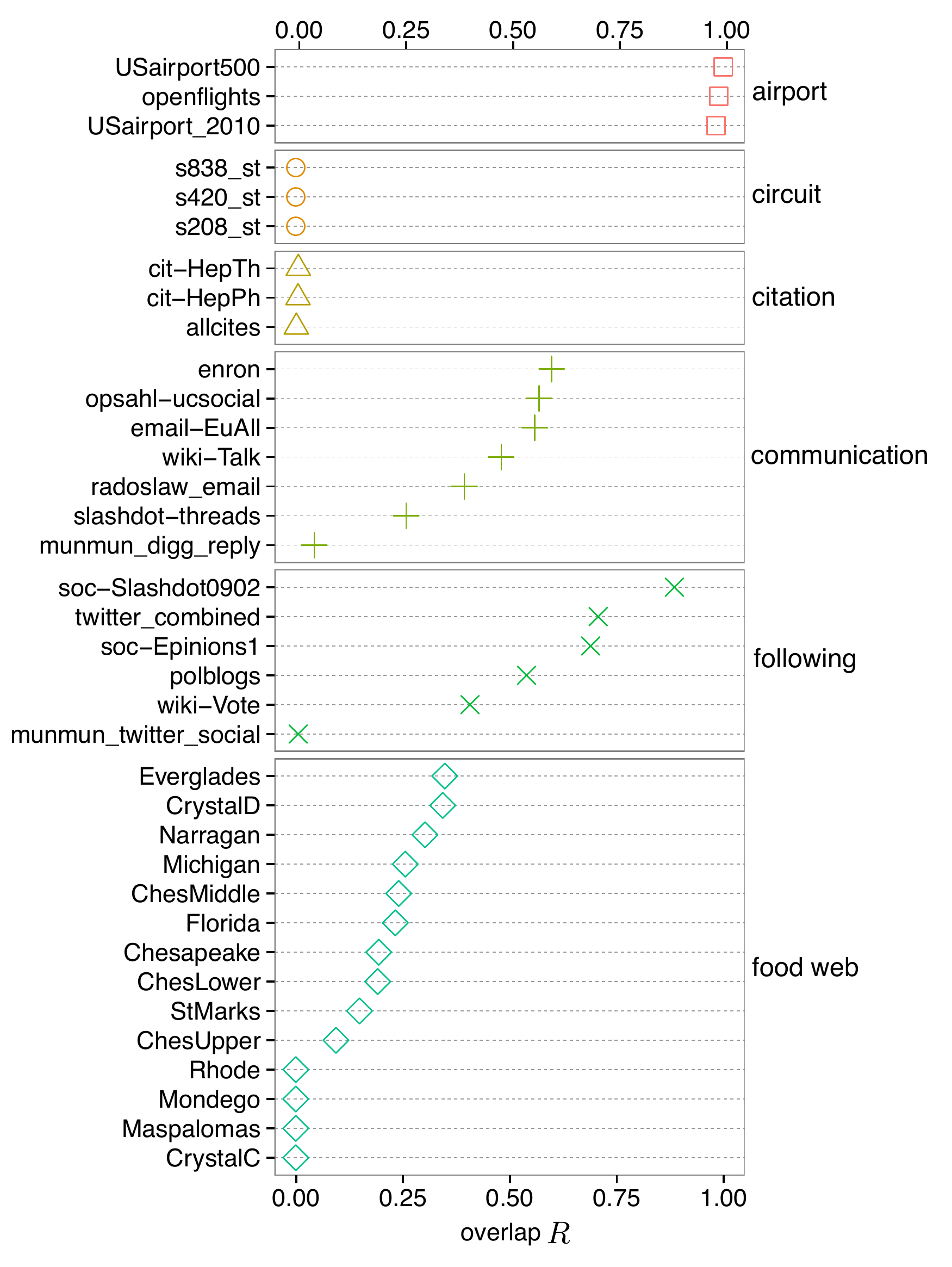}
\caption{
Overlap measure $R$ between the cycle and flow trusses for the airport, circuit, citation, communication, following, and food web networks.}
\label{fig:jaccard1}
\end{figure}

\begin{figure}
\centering
\includegraphics[width=0.8\hsize]{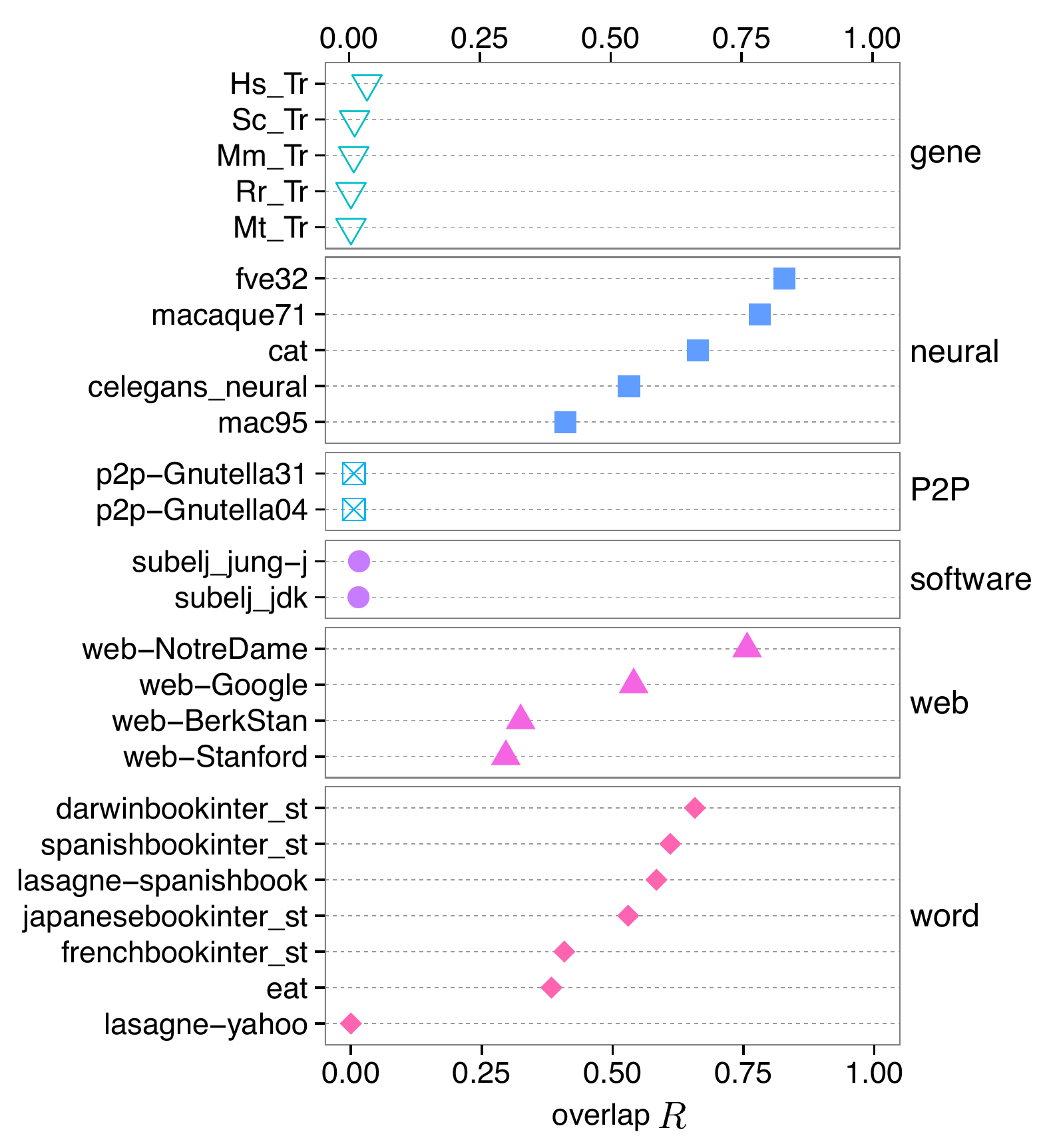}
\caption{
Overlap measure $R$ between the cycle and flow trusses for the gene, neural, P2P, software, web, and word networks.}
\label{fig:jaccard2}
\end{figure}

In~Figs.~\ref{fig:jaccard1} and \ref{fig:jaccard2}, we plot the $R$ values for the empirical networks (i.e., the same set that we used in~\figref{truss_number_D}, except for the metabolic networks).
First, we can see that the networks of some categories have the $R$ values close to the extreme cases, i.e., $0$ or $1$.
For the airport networks, the $R$ values for the three networks are almost equal to unity.
This result follows logically because the reciprocity of links, defined by the double of the number of bidirectionally adjacent node pairs divided by the total number of links, are large: $0.972$, $1$, and $0.781$ for the openflights, USairport500, and USairport\_2010 networks, respectively (see Table~S1 in SM).
Therefore, any triplet of nodes is likely to constitute a cycle triangle if it composes a flow triangle and vice versa.
For the circuit, citation, gene regulatory, P2P, and software networks, the $R$ values are close to zero, because these types of networks have huge gaps between the number of cycle and flow triangles (Tables~S1 and S2 in SM).
The three circuit networks do not contain any flow triangles.
For the citation, gene regulatory, P2P, and software networks, the number of cycle triangles is much smaller than that of flow triangles.

The following, neural, web, and word networks tend to have $R$ values greater than $0.5$, although fluctuations within a category are large.
The food webs tend to have $R$ values less than $0.5$.
The results for the metabolic networks are shown in Fig.~S3 in SM; the 166 out of 172 networks have $R$ values in $\left[ 0.4, 0.6 \right]$ (the mean $R$ value $\pm$ the standard deviation is equal to $0.491 \pm 0.0481$).
These observations may indicate the usefulness of the $R$ value to characterize the tendency of module organization for different categories of networks.

\section{Discussion}
In this paper, we proposed the cycle and flow $k$-trusses in order to extract two distinct types of cohesive modules from directed networks.
We also developed an efficient algorithm for computing these trusses and defined the measures used to quantify the module organization in a network based on the truss properties.
Applications of our method to a wide variety of empirical networks illustrated that most empirical networks contain either type of trusses or both of them.
We investigated the extracted trusses for several networks with the given node labels and found that the trusses seem to capture relevant subgroups of nodes.
We also found that the abundance of (and the overlap between) cycle and flow trusses helps us to classify empirical networks obtained from different fields.
These findings suggest the importance of exploring different types of modules in directed networks.
We believe that our method will be a useful tool for investigating module structure in directed networks.

It is worth noting several relationships betweeen the cycle and flow trusses and other notions of directed subgraphs presented in previous studies.
Detecting directed triangles that are significantly over-presented is a key idea of motif analysis~\cite{Milo2002,Alon2007}.
The original notion of the motif often focuses on subgraphs with a small number of nodes (e.g., three or four).
Previous studies~\cite{Dobrin2004,Kashtan2004,Zhang2005,Michoel2011} considered the generalization of motifs by aggregating the motifs sharing links so as to construct functional modules larger than single motifs.
In particular, the generalization of the feed-forward loop motif (called the flow triangle in this paper) described in Refs.~\cite{Dobrin2004,Kashtan2004,Zhang2005,Michoel2011} is an example of the flow $1$-truss in our definition.
Another related notion is the directed $k$-clique~\cite{Palla2007}.
A directed $k$-clique is a subgraph with $k$ nodes and $k(k-1)/2$ links, in which the $k$ nodes has a linear ordering and all node have directed links to all the lower-rank nodes.
A directed $k$-clique module is defined by a union of adjacent directed $k$-cliques.
Two directed $k$-cliques are said to be adjacent if the two have $k-1$ nodes in common.
A directed $k$-clique module is a flow $(k-2)$-truss $(k \geq 3)$; however, the converse does not always hold true.
Therefore, the cohesiveness of the flow $k$-trusses is between those of the generalization of feed-forward loop motif and the directed $k$-clique modules.
To the best of our knowledge, a cycle $k$-truss does not exactly correspond to any of the previous notions. 
Based on the definition, a cycle truss is a strongly connected component (i.e., any node in the cycle truss is connected to all the other nodes via directed links).

Recently, a graph partitioning method based on the so-called motif conductance is proposed~\cite{Benson2016}. This method focuses on a given motif (e.g., the cycle or flow triangle) and splits a network into two parts so as to minimize the ratio of the number of the motifs crossing the two parts to the number of the motifs contained by either part (taking the minimum value for the two parts). Repetitive application of the method is expected to result in a partition of the network in which each part contains many of the focal motifs.
Although the aim of this method is different from our method, it would be interesting to see the overlap and difference between the results of the two methods. As an example, we applied the motif conductance method\footnote{Codes were downloaded from \url{http://snap.stanford.edu/higher-order/} (Date Accessed: 13th October 2016).} to the C.~elegans neural network (section~3(a)(i)). The method returned four nontrivial components for the cycle triangle and two for the flow triangles. The node set of the four components for the cycle triangle has no intersection with the node set of the cycle $3$-truss (Fig.~\ref{fig:demo_cele}(b)). The node set of a component for the flow truss contains the node set of the flow $9$-truss (Fig.~\ref{fig:demo_cele}(c)). In this sense, the motif conductance method for the flow triangle and the truss method are consistent for this example. The cycle $3$-truss is not captured by the motif conductance method maybe because the method prefers increasing the number of focal motifs inside of the two parts to decreasing the number of motifs crossing the partition. Further analysis on comparison between the two methods is future work.

Our definition of trusses in this paper assumes that the focal network is directed and unweighted.
However, the importance of link weight in many systems has been suggested in previous work~(e.g., Ref.~\cite{Barrat2004}).
This point is a clear limitation of the present method and a suitable generalization for weighted networks is warranted.
Although we determined the existence of truss structures in empirical networks, the origins and functional roles of these trusses are not yet well understood.
Functional roles of the generalized motifs in biological networks have been investigated~\cite{Dobrin2004,Kashtan2004,Zhang2005,Michoel2011}. A similar investigation on the functionality of trusses would be the next step and require the knowledge of field experts and well-documented network data with link annotations such as gene ontology databases.
Finally, dynamical models of the growth processes of directed networks that yield truss structures will be potential future work, providing further understanding of the organization of modules in directed networks.

\section*{Acknowledgments}
The authors thank Naoki Masuda for insightful comments.
\textbf{Author contributions:} TT and YY conceived and designed the research.
YY performed the coding.
TT and YY analyzed the data, discussed the results, and wrote the manuscript.
\textbf{Competing interests:} The authors declare no competing financial interests.
\textbf{Funding:} YY acknowledges the financial support through JST, ERATO, Kawarabayashi Large Graph Project, JSPS Grant-in-Aid for Young Scientists (B) (No.~26730009), and MEXT Grant-in-Aid for Scientific Research on Innovative Areas (No.~24106003).


\begin{thebibliography}{34}

\bibitem{Newman2010}
Newman MEJ. 2010
 {\em {Networks: an Introduction}}.
 Oxford: Oxford University Press.

\bibitem{Costa2007}
Costa LF, Rodrigues FA, Travieso G, {Villas Boas} PR. 2007
 {Characterization of complex networks: A survey of measurements}.
 {\em Adv. Phys.} {\bf 56}, 167--242.

\bibitem{Barrat2008}
Barrat A, Barth{\'{e}}lemy M, Vespignani A. 2008
 {\em {Dynamical Processes on Complex Networks}}.
 Cambridge: Cambridge University Press.

\bibitem{Porter2009}
Porter MA, Onnela J-P, Mucha PJ. 2009
 {Communities in Networks}.
 {\em Not. Am. Math. Soc.} {\bf 56}, 1082--1097.

\bibitem{Fortunato2010}
Fortunato S. 2010
 {Community detection in graphs}.
 {\em Phys. Rep.} {\bf 486}, 75--174.

\bibitem{Rosvall2008}
Rosvall M, Bergstrom CT. 2008
 {Maps of random walks on complex networks reveal community
  structure.}
 {\em Proc. Natl. Acad. Sci. USA} {\bf 105}, 1118--23.

\bibitem{Clauset2008}
Clauset A, Moore C, Newman MEJ. 2008
 {Hierarchical structure and the prediction of missing links in
  networks}.
 {\em Nature} {\bf 453}, 98--101.

\bibitem{Chen2006}
Chen J., Yuan B. 2006
 {Detecting functional modules in the yeast protein-protein
  interaction network}.
 {\em Bioinformatics} {\bf 22}, 2283--2290.

\bibitem{Sohn2011}
Sohn Y, Choi M-K, Ahn Y-Y, Lee J, Jeong J. 2011
 {Topological cluster analysis reveals the systemic organization of
  the Caenorhabditis elegans connectome.}
 {\em PLOS Comput. Biol.} {\bf 7}, e1001139.

\bibitem{Chung2005}
Chung F. 2005
 {Laplacians and the Cheeger Inequality for Directed Graphs}.
 {\em Ann. Comb.} {\bf 9}, 1--19.

\bibitem{Michoel2012}
Michoel T, Nachtergaele B. 2012
{Alignment and integration of complex networks by hypergraph-based spectral clustering}.
{{\em Phys. Rev.} E} {\bf 86}, 056111. 

\bibitem{Yoshida2016}
Yoshida Y. 2016
 {Nonlinear Laplacian for digraphs and its applications to network analysis}.
 In {\em Proc. the Ninth ACM International Conference on Web Search and Data Mining}, San Francisco, CA, USA, 22 to 25 February 2016, pp. 483--492.

\bibitem{Leicht2008}
Leicht EA, Newman MEJ. 2008
 {Community structure in directed networks}.
 {\em Phys. Rev. Lett.} {\bf 100}, 118703.

\bibitem{Malliaros2013}
Malliaros FD, Vazirgiannis M. 2013
 {Clustering and community detection in directed networks: A survey}.
 {\em Phys. Rep.} {\bf 533}, 95--142.

\bibitem{Kim2010}
Kim Y, Son S-W, Jeong H. 2010
 {Finding communities in directed networks}.
 {{\em Phys. Rev.} E} {\bf 81}, 016103.

\bibitem{Broder2000}
Broder A, Kumar R, Maghoul F, Raghavan P, Rajagopalan S, Stata R, Tomkins A, Wiener J. 2000
 {Graph structure in the Web}.
 {\em Comput. Networks} {\bf 33}, 309--320.

\bibitem{Corominas-Murtra2013}
Corominas-Murtra B, Go{\~{n}}i J, Sol{\'{e}} RV, Rodr{\'{i}}guez-Caso C. 2013
 {On the origins of hierarchy in complex networks.}
 {\em Proc. Natl. Acad. Sci. U.S.A.} {\bf 110}, 13316--13321.

\bibitem{Milo2002}
Milo R, Shen-Orr S, Itzkovitz S, Kashtan N, Chklovskii D, Alon U. 2002
 {Network motifs: simple building blocks of complex networks.}
 {\em Science (New York, N.Y.)} {\bf 298}, 824--827.

\bibitem{Alon2007}
Alon U. 2007
 {Network motifs: theory and experimental approaches.}
 {\em Nat. Rev. Genet.} {\bf 8}, 450--461.

\bibitem{Cohen2008}
Cohen J. 2008
 {Trusses : Cohesive Subgraphs for Social Network Analysis}.
 Technical report, National Security Agency, Fort Meade, MD.

\bibitem{note_truss_def}
In the original definition for undirected networks~\cite{Cohen2008}, the $k$-truss is the maximal subgraph in which every link is involved in $(k-2)$ triangles, not $k$ triangles. We modify the definition in order to make the truss number of links involved in no triangles zero.

\bibitem{Wang2012}
Wang J, Cheng J. 2012
 {Truss decomposition in massive networks}.
 {\em Proc. VLDB Endowment} {\bf 5}, 812--823.

\bibitem{Varshney2011}
Varshney LR, Chen BL, Paniagua E, Hall DH, Chklovskii DB. 2011
 {Structural properties of the Caenorhabditis elegans neuronal network}.
 {\em PLOS Comput. Biol.} {\bf 7}, e1001066.

\bibitem{Kiss1973}
Kiss GR, Armstrong C, Milroy R, Piper J. 1973
 {An associative thesaurus of English and its computer analysis}.
 In {\em The computer and literary studies} (eds. Aitken AJ, Bailey RW, Hamilton-Smith N), pp. 153--165. Edinburgh, UK: Edinburgh University Press.
 
\bibitem{Altun2013}
Altun ZF, Hall DH.
 {Nervous system, general description}, In {\em WormAtlas}, \url{http://dx.doi.org/doi:10.3908/wormatlas.1.18} (Date accessed: 1st March 2016).

\bibitem{Newman2001}
Newman MEJ, Strogatz SH, Watts DJ. 2001
 {Random graphs with arbitrary degree distributions and their applications}.
 {{\em Phys. Rev.} E} {\bf 64}, 026118.

\bibitem{Watts1998}
Watts DJ, Strogatz SH. 1998
 {Collective dynamics of `small-world' networks.}
 {\em Nature} {\bf 393}, 440--442.

\bibitem{Opsahl2011}
Opsahl T.
 {Why Anchorage is not (that) important: Binary ties and Sample selection}, \url{http://toreopsahl.com/2011/08/12/why-anchorage-is-not-that-important-binary-ties-and-sample-selection/} (Date accessed: 1st March 2016).

\bibitem{Leskovec2009}
Leskovec J, Lang KJ, Dasgupta A, Mahoney MW. 2009
{Community structure in large networks: Natural cluster sizes and the absence of large well-defined clusters}.
{\em Internet Math.} {\bf 6}, 29--123 (2009).

\bibitem{Dobrin2004}
Dobrin R, Beg QK, Barab{\'{a}}si A-L, Oltvai ZN. 2004
 {Aggregation of topological motifs in the Escherichia coli transcriptional regulatory network.}
 {\em BMC Bioinform.} {\bf 5}, 10.

\bibitem{Kashtan2004}
Kashtan N, Itzkovitz S, Milo R, Alon U. 2004
 {Topological generalizations of network motifs}.
 {{\em Phys. Rev.} E} {\bf 70}, 031909.
 
\bibitem{Zhang2005}
Zhang LV, King OD, Wong SL, Goldberg DS, Tong AHY, Lesage G, Andrews B, Bussey H, Boone C, Roth FP. 2005
 {Motifs, themes and thematic maps of an integrated Saccharomyces cerevisiae interaction network.}
 {\em J. Biol.} {\bf 4}, 6.

\bibitem{Michoel2011}
Michoel T, Joshi A, Nachtergaele B, {Van de Peer} Y. 2011
 {Enrichment and aggregation of topological motifs are independent organizational principles of integrated interaction networks.}
 {\em Mol. Biosyst.} {\bf 7}, 2769--2778.

\bibitem{Palla2007}
Palla G, Farkas IJ, Pollner P, Der{\'{e}}nyi I, Vicsek T. 2007
 {Directed network modules}.
 {\em New J. Phys.} {\bf 9}, 186.

\bibitem{Benson2016}
Benson AR, Gleich DF, Leskovec J. 2016
{Higher-order organization of complex networks}.
{\em Science} {\bf 353}, 163--166.
 
\bibitem{Barrat2004}
Barrat A, Barth{\'{e}}lemy M, Pastor-Satorras R, Vespignani A. 2004
 {The architecture of complex weighted networks}.
 {\em Proc. Natl. Acad. Sci. U.S.A.} {\bf 101}, 3747--3752.
 
\end{thebibliography}

\clearpage

\clearpage

\clearpage

\clearpage

\clearpage

\clearpage

\end{document}